\documentclass[aps,prl,twocolumn,showpacs,psfig,superscriptaddress,longbibliography]{revtex4-1}

\usepackage{textcomp}
\usepackage{times}
\usepackage{graphicx}
\usepackage{float}
\usepackage{latexsym,amsmath,amssymb,bm,euscript}
\usepackage{color}
\usepackage{subfigure}
\usepackage{epstopdf}
\usepackage[colorlinks=true,linkcolor=blue,citecolor=blue]{hyperref}
\usepackage{hyperref}
\usepackage{soul}
\usepackage[normalem]{ulem}
\usepackage{mathrsfs}
\usepackage{amsmath}
\usepackage{lettrine}
\usepackage{xspace}
\usepackage{textcomp}

\newcommand{\wzl}[1]{\textcolor{black}{#1}}

\begin{document}
\title{NMR evidence of pressure-induced structural transition and enhanced spin fluctuations \wzl{up to 14~GPa} in SrCu$_2$(BO$_3$)$_2$ }

\author{Zhanlong Wu}
\thanks{These authors contributed equally to this study.}
\affiliation{School of Physics and Key Laboratory of Quantum State Construction and Manipulation (Ministry of Education), Renmin University of China, Beijing 100872, China}

\author{Kefan Du}
\thanks{These authors contributed equally to this study.}
\affiliation{School of Physics and Key Laboratory of Quantum State Construction and Manipulation (Ministry of Education), Renmin University of China, Beijing 100872, China}

\author{Shuo Li}
\thanks{These authors contributed equally to this study.}
\affiliation{Institute of Physics, Chinese Academy of Sciences, and Beijing National Laboratory for Condensed Matter Physics, Beijing 100190, China}

\author{Tong Shi}
\thanks{These authors contributed equally to this study.}
\affiliation{Institute of Physics, Chinese Academy of Sciences, and Beijing National Laboratory for Condensed Matter Physics, Beijing 100190, China}

\author{Ying Chen}
\thanks{These authors contributed equally to this study.}
\affiliation{School of Physics and Key Laboratory of Quantum State Construction and Manipulation (Ministry of Education), Renmin University of China, Beijing 100872, China}

\author{Qingxin Dong}
\affiliation{Institute of Physics, Chinese Academy of Sciences, and Beijing National Laboratory for Condensed Matter Physics, Beijing 100190, China}

\author{Rui Zhou}
\affiliation{Institute of Physics, Chinese Academy of Sciences, and Beijing National Laboratory for Condensed Matter Physics, Beijing 100190, China}

\author{Rong Yu}
\affiliation{School of Physics and Key Laboratory of Quantum State Construction and Manipulation (Ministry of Education), Renmin University of China, Beijing 100872, China}


\author{Juanjuan Liu}
\email{juanjuanliu@ruc.edu.cn}
\affiliation{School of Physics and Key Laboratory of Quantum State Construction and Manipulation (Ministry of Education), Renmin University of China, Beijing 100872, China}

\author{Bosen Wang}
\email{bswang@iphy.ac.cn}
\affiliation{Institute of Physics, Chinese Academy of Sciences, and Beijing National Laboratory for Condensed Matter Physics, Beijing 100190, China}

\author{Jinguang Cheng}
\email{jgcheng@iphy.ac.cn}
\affiliation{Institute of Physics, Chinese Academy of Sciences, and Beijing National Laboratory for Condensed Matter Physics, Beijing 100190, China}

\author{Weiqiang Yu}
\email{wqyu\_phy@ruc.edu.cn}
\affiliation{School of Physics and Key Laboratory of Quantum State Construction and Manipulation (Ministry of Education), Renmin University of China, Beijing 100872, China}

\author{Yi Cui}
\email{cuiyi@ruc.edu.cn}
\affiliation{School of Physics and Key Laboratory of Quantum State Construction and Manipulation (Ministry of Education), Renmin University of China, Beijing 100872, China}


\begin{abstract}

\wzl{The Shastry-Sutherland compound SrCu$_2$(BO$_3$)$_2$ has attracted considerable interest as a platform for exploring quantum phases and quantum phase transitions driven by magnetic frustration. The pressure-induced structural and magnetic phase transitions in SrCu$_2$(BO$_3$)$_2$, however, remain controversial. To address this issue, we performed high-pressure $^{11}$B nuclear magnetic resonance (NMR) measurements on SrCu$_2$(BO$_3$)$_2$ up to 14~GPa. The NMR spectra reveal two pressure-induced monoclinic phases. With pressure above 4~GPa and with temperature below 10~K, the rapid broadening of the NMR spectrum and the power-law behavior in the spin-lattice relaxation rate $1/T_1$ provide clear evidence for a gapless 3D antiferromagnetic (AFM) phase in the monoclinic phase. At an intermediate temperature range around 20~K, the emergence of the field-dependent NMR line splits resolves a two-dimensional, short-range ordered AFM phase; at temperature above 30~K, the sublinear power-law behavior of $1/T_1$ identifies an extended correlated paramagnetic regime.}

\end{abstract}

\maketitle

{\bf Introduction.}  
Frustrated magnetic materials provide an excellent platform for studying novel phase transitions and exotic states, including quantum spin liquids, resonating valence bond states, and spin ice~\cite{Balents2010Nature,Savary_2017,RevModPhys.89.025003}. A paradigmatic example is the two-dimensional (2D) Shastry-Sutherland model (SSM), characterized by antiferromagnetic intradimer and interdimer exchange couplings, denoted by $J'$ and $J$, respectively~\cite{SS1981}. Its ground-state phase diagram is controlled by the coupling ratio $\alpha=J/J'$. For $\alpha\lesssim0.675$, the ground state is an exact product of dimer singlets~(DS)~\cite{SS1981,Koga2000PRL}. With increasing $\alpha$, the system undergoes a transition into a plaquette-singlet~(PS) phase for approximately $0.675\lesssim\alpha\lesssim0.765$~\cite{Koga2000PRL,Corboz2013PRB}, followed by an antiferromagnetic (AFM) order phase at larger $\alpha$~\cite{Manousakis1991,2018PRX}.

Recently, it is found that SrCu$_2$(BO$_3$)$_2$ is a spin-1/2 2D magnetic material that closely realizes the Shastry-Sutherland model~\cite{Kageyama_1999_JPSJ}. At ambient pressure, it lies in the DS phase, with $J'$ $\approx$ 85~K and $J$ $\approx$ 53~K ($\alpha$ $\approx$ 0.63)~\cite{Miyahara1999PRL}. Under applied pressure, it transitions successively into the PS phase ($\sim$ 1.8~GPa) and the SSM-AFM phase ($\sim$ 2.78~GPa)~\cite{JingGuo2020PRL,JingGuo2025}. Under an external magnetic field, it also exhibits a rich variety of quantum states, including fractional magnetization plateaus~\cite{Kageyama_1999_JPSJ,K.Onizuka2000JPSJ,S.Miyahara2000PRB,WOLF20011973,Y.Fukumoto2001JPSJ,D.A.Schneider2016PRB}, spin-nematic phase~\cite{S.C.Furuya2018PRB,S.Imajo2022PRL,Fogh2024}, spin-supersolid state~\cite{T.Momoi2000PRB,P.Corboz2014PRL,Shi2022NC,T.Nomura2023NC}, and a field-induced proximate deconfined quantum critical point~\cite{Cui2023,Cui_2025}.

\wzl{Previous high-pressure synchrotron X-ray diffraction and Raman studies revealed a weak tetragonal-to-monoclinic structural distortion in SrCu$_2$(BO$_3$)$_2$ near 4.7~GPa at room temperature ~\cite{Loa2005}. This distortion was later followed over the temperature-pressure phase diagram by angle-dispersive synchrotron powder and single-crystal X-ray diffraction, which placed its onset in the narrow range of 4.0–5.0~GPa~\cite{Zayed2014} (seen in Fig.~\ref{Fig4}). However, with increasing pressure, the N\'eel temperature $T_{\rm N}$ shows a pronounced discontinuous increase already around 4~GPa.} Unlike the tetragonal phase which exhibits only one AFM transition, two possible magnetic transitions were proposed at 125~K and 8~K, respectively, in the monoclinic phase~\cite{JingGuo2020PRL}. However, neutron diffraction detected an enhancement of magnetic Bragg peaks around 120~K, without a second transition at low temperatures~\cite{S.Haravifard2014PNAS,Zayed2017NaturePhysics, FoghEllen2024PRL}. At 5.5~GPa, inelastic neutron scattering (INS) measured at 4.5~K revealed that the intra-plane exchange interactions remain dominant with $J'$ $\approx$ 27~K and $J$ $\approx$ 49~K, while the interlayer coupling is much weaker, $J_c\approx0.6$ K~\cite{FoghEllen2024PRL}. Although  $J_{\rm c}$ is weak, it lifts the degeneracy of the two Goldstone branches, rendering one branch gapped while the other remains gapless. Owing to the AFM interlayer stacking, the corresponding spectral weight is absent at $\mathbf{Q}=(0,1,0)$ and $(0,3,0)$, while the mode at $(1,1,0)$ is also not confirmed by the INS data due to low-energy limitations.

\begin{figure*}[t]
    \includegraphics[width=18cm]{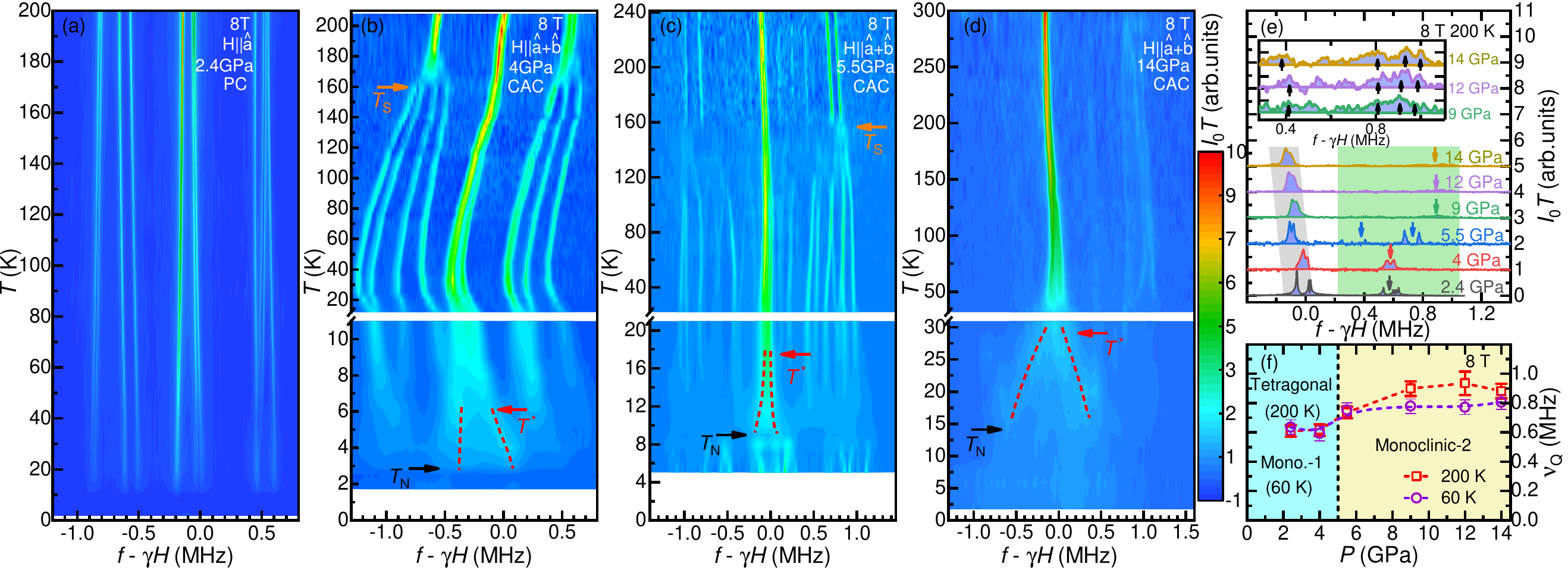}
    \centering
    \caption{\label{Fig1}
    {\bf NMR spectra.} 
    (a) - (d) Colour map of NMR spectra at 2.4~GPa, 4~GPa, 5.5~GPa and 14~GPa, 8~T. $T_{\rm S}$ (orange arrows) is the structural transition temperature. $T_{\rm N}$ (black arrows) is the N\'eel temperature of the 3D AFM order. $T^*$ (red arrows) is the onset temperature of AFL state. The separation of the NMR lines is marked by the red dashed lines in (b) - (d).
    (e) NMR spectra at 200~K under typical pressures (data shifted vertically for clarity). The gray area marks the central peaks and  the green area marks the satellite peaks on the right. The arrows mark the positions of the satellite peaks. \wzl{Inset: an enlarged view of the satellite peaks between 9 and 14 GPa, where the four‑peak structure is noted by black arrows.}
    (f) Quadrupole frequency $\nu_{\rm Q}$  extracted from (e) as functions of pressures. The vertical dashed line denotes the structural boundaries separating the tetragonal and monoclinic-2 phases at~200 K, and the monoclinic-1 and monoclinic-2 phases at 60~K, respectively.
}
\end{figure*}

To resolve these issues in a low-energy aspect, we performed high-pressure NMR measurements on SrCu$_2$(BO$_3$)$_2$ with up to 14~GPa and down to 1.8~K, extending NMR spectroscopy into a pressure regime that remains largely unexplored in strongly correlated quantum materials. \wzl{A marked reconstruction of the satellite spectra near 4 GPa signals a transition from the tetragonal phase to a monoclinic-1 phase. The pressure evolution of the quadrupolar frequency $\nu_{\rm Q}$ further reveals a second structural transition, from monoclinic-1 to monoclinic-2, near 5~GPa} In the monoclinic phases, pronounced line broadening together with peaks in the spin-lattice relaxation rate $1/T_1$ identifies AFM transitions at approximately 10~K at 5.5~GPa and 15~K at 14~GPa. The $1/T_1$ exhibits a power-law temperature dependence below the transition, which is consistent with a gapless phase. Above the transition, a line separation is found, which resolves a short-range ordered, antiferromagnetic liquid state. \wzl{At still higher temperatures, extending to approximately 80 K, the Knight shift develops a broad maximum, accompanied by a power-law dependence of $1/T_1$, marking the onset of a correlated paramagnetic regime.}
    
{\bf Experimental details.} 
High-quality single crystals of SrCu$_2$(BO$_3$)$_2$ were grown by the optical floating zone method~\cite{KAGEYAMA199965}. Pressures up to 2.4~GPa are achieved by a piston cell (PC) with Daphne 7373~oil as the pressure-transmitting medium. Pressures from 2.65~GPa to 14~GPa are achieved in a cubic anvil cell (CAC), with glycerol as the pressure-transmitting medium. \wzl{The utilization of these two types of pressure cells optimize the pressure-hydrostaticity to our knowledge.} 

 We performed $^{11}$B NMR measurements ($I=3/2$, $\gamma$~=~13.655~MHz/T) with the magnetic field applied within the $ab$ plane. \wzl{With the magnetic field applied in the $ab$ plane, the four inequivalent boron sites in SrCu$_2$(BO$_3$)$_2$ allow us to sensitively monitor the splitting of quadrupolar satellite lines, which reflect the lowering of structural symmetry~\cite{Waki2007}. Because SrCu$_2$(BO$_3$)$_2$ is characterized by nearly isotropic Heisenberg exchange interactions and a weakly anisotropic $g$ factor, substantial field-orientation effects are not expected.}  

$^{11}$B NMR spectra were collected by the spin-echo method, by sweeping frequencies in a fixed field. The NMR Knight shift $K_n$ was calculated by $K_n=[\left<f\right>/\gamma H-1]~\times~100\% $, where $\left<f\right>$ represents the average frequency of the central peaks. $1/T_1$ was measured by the spin inversion-recovery method \wzl{on the $^{11}$B central peaks } with the nuclear magnetization $M(t)$ fitted to the recovery function $M(t)=M(\infty)[1-be^{-(t/T_1)^{\beta}}-9be^{-(6t/T_1)^{\beta}}]$, \wzl{such multi‑exponential recovery arising from multi‑level relaxation for spin-3/2 nuclear spins~\cite{Albert1967,Chepin_1991}. $\beta$ is a stretching exponent and is found to be about 1 in the paramagnetic phase for both the PC and CAC measurements, indicating a nearly single-component spin-lattice relaxation. This supports the high quality of the crystals and the high pressure hydrostaticity of the pressure medium.}
   
{\bf Structural phase transitions.} 
To clarify the origin of the high-temperature anomaly of specific heat and magnetic Bragg peak under high pressure, we performed high-pressure NMR spectral measurements. Figures~\ref{Fig1} (a) - (d) show the NMR spectra at 2.4~GPa, 4~GPa, 5.5~GPa and 14~GPa, respectively, under a magnetic field of 8~T. A central peak and two satellite peaks can be clearly resolved as expected for spin-3/2 nuclei. As shown in Fig.~\ref{Fig1}(a), with external magnetic field aligned along the $a$-axis, four inequivalent $^{11}$B sites in SrCu$_2$(BO$_3$)$_2$ produce four distinct peaks for both the center and satellite lines. At 2.4~GPa, the NMR spectra exhibit no line splitting with temperature down to 1.8~K.

At 4~GPa [Fig.~\ref{Fig1}(b)], all central and satellite peaks above 180~K appear as single-line structures, as the sample was placed with field along the direction of [110]~\cite{Waki2007}. \wzl{Below 160~K, the previously overlapping satellite peaks from the four inequivalent $^{11}$B sites become clearly separated, whereas the central peak remains almost unchanged. Therefore, this behavior indicates a modification of the local electric field gradients (EFGs) associated with a structural transition. Consistently, no magnetic anomaly is detected in the spin-lattice relaxation rate over the same temperature range (see Fig. \ref{Fig3}), arguing against a magnetic origin of the transition. We therefore identify this transition as the structural transition from the tetragonal to the monoclinic-1 phase and denote the transition temperature by $T_{\rm S}$. These line splits are inconsistent with short-range inhomogeneous lattice or electronic correlations, which would produce a distribution of local EFGs and hence broadened or asymmetric lines.}

The structural phase transition is also resolved by the satellite peaks at 5.5~GPa, with $T_{\rm S} \sim 160$~K. However, above $T_{\rm S}$, the number of resolved satellite lines is approximately doubled. We propose that this is caused by the fact that 5.5~GPa is close to the structural transition from the tetragonal to the monoclinic-2 phase, leading to the phase separation of two phases. At 14~GPa, four satellite peaks are seen again \wzl{[inset of Fig. \ref{Fig1}(e)]}, indicating that SrCu$_2$(BO$_3$)$_2$ is in a homogeneous monoclinic phase at this pressure.

  \begin{figure}[t]
    \centering
    \includegraphics[width=8.5cm]{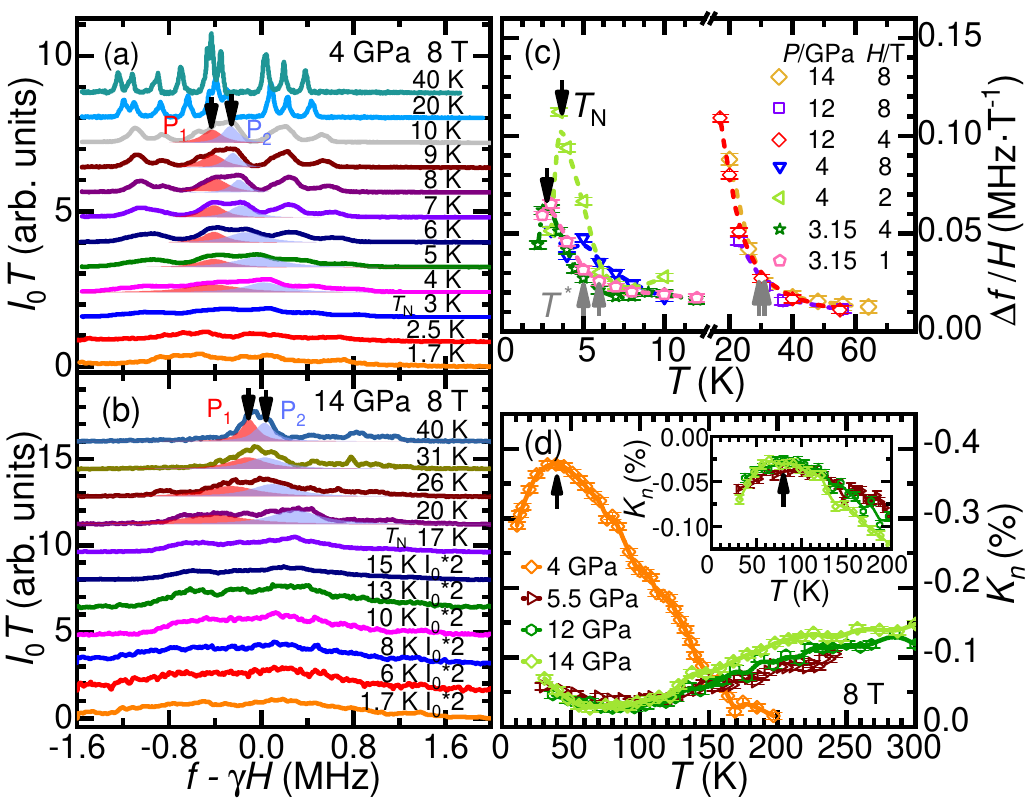}
    \caption{\label{Fig2}
    {\bf Low-temperature NMR spectra.}
    (a)-(b) Low temperature NMR spectra at 4~GPa and 14~GPa, respectively, measured under a field of 8~T applied in the $ab$ plane. The black arrows mark the \wzl{line separation} of central peaks. $T_{\rm N}$ is the N\'{e}el temperature of the 3D AFM transition. \wzl{The double Lorentzian fit to the central peak is depicted by red and light‑blue shaded areas.}
    (c) \wzl{Line separation of central peak, $\Delta f$ = $|f_2 - f_1|$, at different pressures and scaled by magnetic field. $f_1$ and $f_2$ are frequencies of $\rm {P_1}$ and $\rm {P_2}$ respectively.} $T^*$ marks the onset temperature of the AFL state. 
    (d) Knight shift $K_n$ as functions of temperatures under different pressures at 8~T. The black arrows mark the hump of $K_n$. The structural transition is accompanied by a sign change of the hyperfine coupling constant $A_{\rm hf}$ from the monoclinic-1 phase (as shown at 4~GPa) and the monoclinic-2 phase (as shown at 5.5~GPa and above). Inset: Enlarged view of $K_n$ above 5.5~GPa with an inverted vertical axis scale.
}
\end{figure}

   Figure~\ref{Fig1}(e) shows the NMR spectra at 200~K and 8~T under different pressures. The gray and green shaded regions trace the pressure evolution of the central line and one representative satellite line, respectively. We calculated the average frequencies of the central peak $\left<f\right>_C$ and satellite peak $\left<f\right>_S$ respectively. By this, the values of quadrupole frequency $\nu_{\rm Q}$ are calculated using $\nu_{\rm Q}$ = $\left<f\right>_S -  \left<f\right>_C$, and shown in Fig.~\ref{Fig1}(f). At 200~K, $\nu_{\rm Q}$ is approximately 0.6~MHz at 2.4~and 4~GPa, and then increased to a plateau at about 0.89~MHz with pressure above 9~GPa. The sharp increase indicates the presence of a structural phase transition in the system. A similar pressure evolution of $\nu_{\rm Q}$ is observed at 60~K. Above 5~GPa, a stable monoclinic phase is resolved by the nearly constant value of $\nu_{\rm Q}$. \wzl{In contrast, between approximately 4 and 5~GPa, $\nu_{\rm Q}$ changes rapidly with pressure, suggesting an intermediate structural regime. We therefore identify this pressure range as the monoclinic-1 phase and assign the higher-pressure phase as monoclinic-2 (Fig.~\ref{Fig4}).}  By contrast, previous studies suggest that the system undergoes a single structural phase transition from a tetragonal phase to a monoclinic phase at a lower pressure of 4.7~GPa~\cite{Loa2005,Zayed2014,Zayed2017NaturePhysics}.

\begin{figure}[t]
    \centering
    \includegraphics[width=8.5cm]{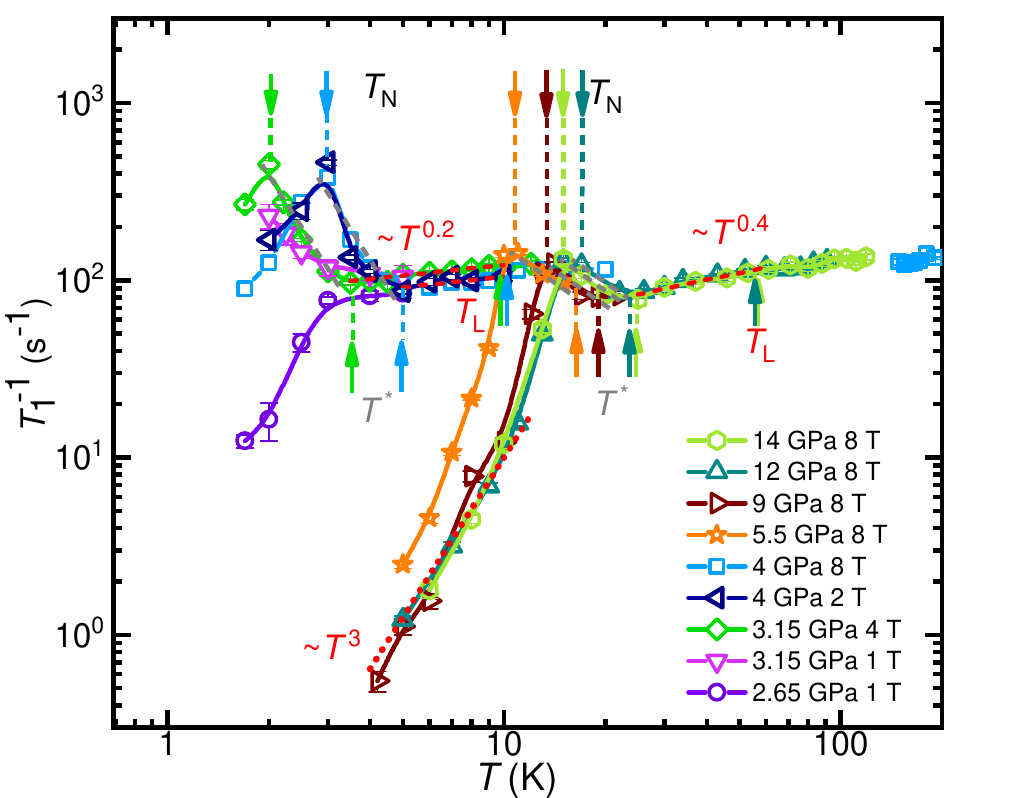}
    \caption{\label{Fig3}
    {\bf The spin-lattice relaxation rates.}
    $1/T_1$ as functions of temperatures under different magnetic fields measured from 2.65~GPa to 14~GPa. $T^*$ marks the onset temperature of the AFL state. $T_{\rm N}$ is the N\'{e}el temperature of the 3D AFM order. $T_{\rm L}$ marks the crossover temperature of the correlated paramagnet. Red dashed and dotted lines represent power-law fits of $1/T_1$. Gray dashed lines indicate the rise in $1/T_1$ due to enhanced spin fluctuations in the AFL state.
}
\end{figure}

    \begin{figure}[t]
    \centering
    \includegraphics[width=8.5cm]{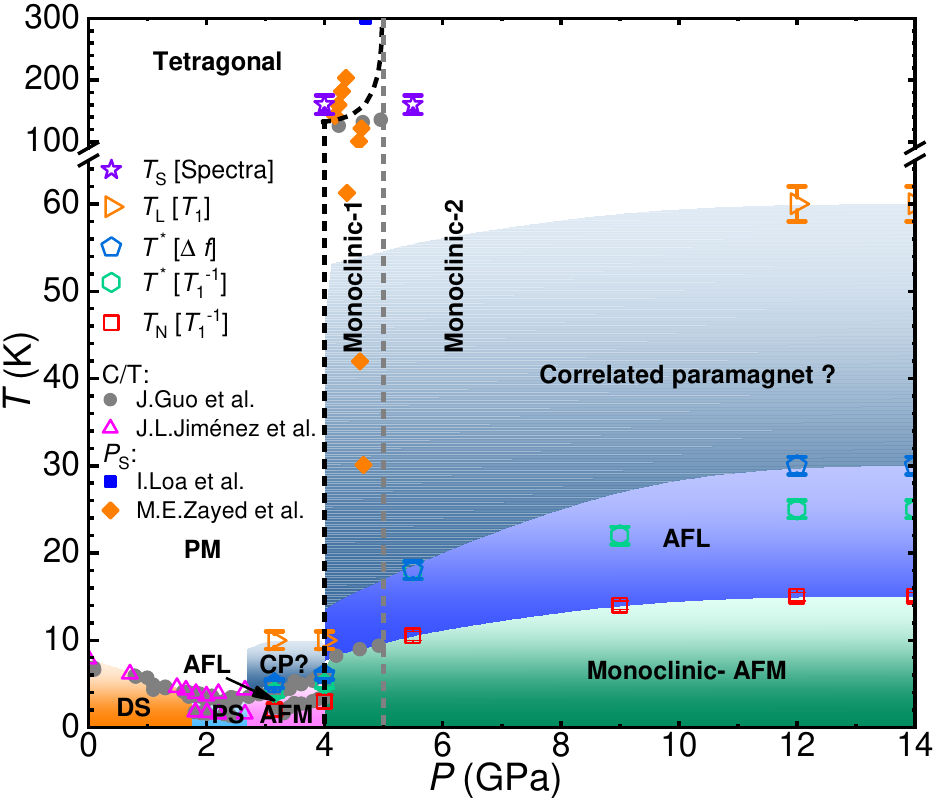}
    \caption{\label{Fig4}
    {\bf \textbf{($P$,$T$)} phase diagram of SrCu$_2$(BO$_3$)$_2$.}
    Tetragonal, monoclinic-1 and monoclinic-2 phases denote three structural phases. The low-temperature ground-state regimes, including DS, PS, tetragonal AFM, and monoclinic AFM, are indicated by different colored regions. The gray dots and magenta triangles represent data from previous specific-heat~\cite{JingGuo2020PRL,Jimenez2021}, \wzl{solid square and diamonds represent structural transition pressures $P_{\rm S}$ from previous work~\cite{Loa2005,Zayed2014}} and others are from our NMR measurements. $T_{\rm S}$ is the symmetry change temperature and $T_{\rm L}$ is the crossover temperature of the proposed correlated paramagnet. $T^*$ is the crossover temperature of AFL state, and $T_{\rm N}$ is N\'{e}el temperature. 
}
\end{figure}

{\bf The AFL state and 3D AFM order.} 
   Figures~\ref{Fig2}(a) and (b) show the low-temperature NMR spectra measured at 4~GPa and 14~GPa, under an 8~T magnetic field. From 10~K to 4~K, \wzl{the central lines gradually broaden and separate [marked by black arrows in Fig.~\ref{Fig2}(a)], and the frequency difference $\Delta f$ between the two peaks ($\rm P_1$ and $\rm P_2$) increases upon cooling, reaching a maximum at $T_{\rm N}$. Notably, $\Delta f$ scales linearly with the applied magnetic field, and the normalized $\Delta f/H$ is plotted in Fig.~\ref{Fig2}(c). The gradual increase and strong field dependence of the line separation above $T_{\rm N}$ are consistent with short-range AFM order, a regime termed the antiferromagnetic liquid (AFL) state in previous studies of SrCu$_2$(BO$_3$)$_2$~\cite{JingGuo2020PRL,JingGuo2025}.} The onset temperature of this short-range order is marked as $T^*$ [Fig.~\ref{Fig2}(c)].

   As shown in Fig.~\ref{Fig2}(b), the characteristic \wzl{line separation associated with} the AFL state persists at 14~GPa and 8~T over the temperature range of 20--40~K. The \wzl{$\Delta f$} remains proportional to the magnetic field [Fig.~\ref{Fig2}(c)], indicating that the AFL state survives across the structural phase transition and remain robust in the monoclinic-2 phase. Below 15~K, the system enters a 3D AFM order state, accompanied by diffusive spectral broadening.

   Figure~\ref{Fig3} shows the temperature dependence of the spin-lattice relaxation rate $1/T_1$ at selected pressures from 2.65~GPa to 14~GPa. At 3.15~GPa and 4~GPa, $1/T_1$ exhibits a low-temperature peak (marked as $T_{\rm N}$), corresponding to the 3D AFM phase transition. The obtained $T_{\rm N}$ values ($\sim$2~K at 3.15~GPa and $\sim$3~K at 4~GPa) agree well with previous zero-field specific heat measurements~\cite{JingGuo2020PRL}. Upon entering the monoclinic phase, the N\'eel temperature shows a discontinuous jump from $\sim$3~K at 4~GPa to 10.5~K at 5.5~GPa. Beyond this, $T_{\rm N}$ increases gradually with pressure and saturates at about 15~K under 14~GPa. Below $T_{\rm N}$ in the monoclinic phase, $1/T_1$ follows a power-law dependence, $\sim T^3$ (red dotted line in Fig.~\ref{Fig3}), revealing gapless excitation modes of the 3D AFM order. This behavior is consistent with the latest INS results ~\cite{Cederholm2026}. As shown in Figs.~\ref{Fig2}(a) and (b), the 3D AFM phase transitions are also revealed by strong spectral broadening and signal loss with the transition temperature denoted as $T_{\rm N}$. The onset temperature of the AFL state is also denoted as $T^*$ in Fig.~\ref{Fig3}. Below $T^*$, the system enters a 2D short-range ordered state, while interlayer magnetic correlations gradually increase and enhanced magnetic fluctuations raise $1/T_1$ (gray dashed lines in Fig.~\ref{Fig3}), eventually driving a 3D long-range AFM ordering at $T_{\rm N}$.

{\bf High-temperature correlated paramagnetism.}
   Figure~\ref{Fig2}(d) shows the temperature dependence of the Knight shift $K_{\rm n}$ at 8~T for pressures from 4 to 14~GPa. A broad maximum is observed near 40~K at 4~GPa and shifts to about 80~K for pressures above 5.5~GPa [marked by black arrows in the enlarged view of Fig.~\ref{Fig2}(d)]. For the 2D Heisenberg model on a square lattice, the magnetic susceptibility exhibits a broad peak around $T \sim J$~\cite{Gomez-Santos_1989_PRB,Hong-Qiang_1991_PRB,Kim_1998_PRL}. Such a broad maximum in the uniform susceptibility is generally associated with the development of short-range AFM correlations, and this behavior suggests that the dominant magnetic correlation energy scale changes substantially across the structural transition.

   In the monoclinic-2 phase, upon cooling below 60~K, $1/T_1$ first decreases gradually (red dashed line in Fig.~\ref{Fig3}) before entering the 3D AFM ordered state. Between 60 and 25~K, $1/T_1$ follows a weak power-law temperature dependence, $\sim T^{0.4}$. A similar power-law behavior, albeit with a lower exponent $\sim T^{0.2}$, is also observed in the tetragonal and monoclinic-1 phases below about 15~K. Here $T_{\rm L}$ is defined phenomenologically as the temperature below which $1/T_1$ deviates from its high-temperature behavior and enters this weak power-law-like regime. At these pressures, the weak power-law \wzl{behavior is observed at temperatures below the broad maximum in the Knight shift, which reflects the characteristic magnetic exchange energy scale}; meanwhile, the NMR spectra exhibit no resolved splitting or static broadening (seen in Fig.~\ref{Fig1}), indicating the absence of static long-range order. 
   
   \wzl{In contrast to the nearly temperature-independent $1/T_1$ expected for a conventional paramagnet, the observed sublinear power-law dependence of $1/T_1$ indicates enhanced low-energy spin fluctuations driven by short-range spin correlations, consistent with a correlated-paramagnetic regime~\cite{Lee2016PRB,Knafo2017PRB,Posp2018PRB,Hovancik2022JSSCh,Graham2023PRL,Chaix2026GeFe2O4}, as also reported in other NMR studies~\cite{Khuntia2016PRL}. Notably, this correlated paramagnetic regime emerges at temperatures well above both the AFL regime and the subsequent long-range AFM ordered state.} 
   A correlated paramagnet represents an intermediate fluctuating regime between a conventional paramagnet and a long-range ordered state, in which low-energy spin fluctuations develop without establishing symmetry-breaking long-range order~\cite{Balents2010Nature,Savary_2017,Cava2010PRB,Ruiz2017NC,Singh_2019}.

{\bf Phase diagram and discussion.}  
Finally, the phase diagram is established as shown in Fig.~\ref{Fig4}.  \wzl{Note that, the phase diagram constructed from our 8~T NMR data provides a reasonable proxy for the zero-field phase diagram above 4~GPa as discussed below. For pressures below 4~GPa, we adopt the previously established zero-field phase boundaries. This consideration is necessary because magnetic fields of approximately 8~T can induce a field-induced PS-to-AFM transition in SrCu$_2$(BO$_3$)$_2$~\cite{Cui2023}, leading to a possible modification of the low-pressure phase boundaries.}

High-pressure drives the system to four ground states sequentially: the DS, PS, tetragonal AFM, and monoclinic AFM phases. The N\'eel temperature $T_{\rm N}$ increases monotonically with pressure, exhibits a discontinuous jump above 4~GPa, and saturates near 12~GPa. Above $T_{\rm N}$, both tetragonal and monoclinic phases host 2D short-range order, as well as a correlated paramagnetic regime at higher temperatures. The black dashed line marks the inferred boundary between the tetragonal and monoclinic phases, while the gray line denotes the boundary between monoclinic-1 and monoclinic-2 phases. 

\wzl{The solid square and diamonds in Fig.~\ref{Fig4}, located in the 4~GPa to 5~GPa range, denote Across this pressure regime, the hyperfine coupling constant $A_{\rm hf}$ changes sign, coinciding with the structural reconstruction from the monoclinic-1 to the monoclinic-2 phase. More importantly, our NMR measurements resolve an intermediate monoclinic-1 regime that was not distinguished in previous bulk structural studies. This additional structural regime provides a natural explanation for the abrupt enhancement of $T_{\rm N}$ near 4~GPa. This work therefore offers a high-resolution local-probe refinement of the high-pressure structural evolution with two distinct monoclinic phases.}

We now discuss the evolution of the exchange couplings. At ambient pressure, SrCu$_2$(BO$_3$)$_2$ exhibits intradimer and interdimer interactions $J' \sim 85$~K and $J \sim 53$~K, respectively. With increasing pressure, both couplings decrease, but $J'$ decreases faster~\cite{Zayed2017NaturePhysics,Sakurai2018}. The broad peak in the magnetic susceptibility at low pressure shifts to lower temperatures, reflecting the reduction of the dimer gap~\cite{Haravifard2012PNAS,Zayed2017NaturePhysics}. In contrast, in the monoclinic phases, the broad peak in the Knight shift appears at higher temperatures: $\sim$40~K in monoclinic-1 and $\sim$80~K in monoclinic-2. This behavior indicates that the dominant magnetic correlation scale is enhanced in the pressure-induced AFM phases and that interdimer interactions become increasingly important. In particular, $J$ in the monoclinic-2 phase is likely much larger than that in monoclinic-1. The jump in $T_{\rm N}$ at 4~GPa suggests that interplane interactions are stronger in the monoclinic phases than in the tetragonal phase, due to the reduced interlayer spacing with pressure~\cite{Loa2005}.

The correlated paramagnetic regime may also be discussed in connection with theoretical proposals of quantum-disordered or gapless quantum spin liquid regimes in the Shastry–Sutherland model between the DS and AFM phases~\cite{Sandvik2022CPL,Wangling2022PRB,XiNing2023PRB,Viteritti2025PRB,Mila2026PRL,Mengziyang2026arxiv,Guo2026arxiv}. The weak power-law behavior of $1/T_1$ and the absence of resolved static spectral splitting suggest persistent low-energy spin fluctuations above the AFL and 3D AFM regimes. These features are qualitatively compatible with an algebraic paramagnetic liquid like finite-temperature scenario~\cite{Chen2018ScienceBulletin}. This phase emerges for $T < J$: while paramagnetic, it exhibits the same algebraic excitation modes as a zero-temperature quantum spin liquid. Moreover, it persists under an applied magnetic field and can even survive above the magnetic ordering temperature~\cite{Chen2018ScienceBulletin,17OPRL}.

\wzl{In contrast with previous reports, our NMR measurements detect only a single 3D AFM phase transition at low temperature around 10~K in the monoclinic phase. As shown in Fig.~\ref{Fig4}, in both tetragonal and monoclinic phases, the characteristic temperature $T^*$ and $T_{\rm N}$ extracted from our finite-field NMR data are consistent with those determined from zero-field specific heat measurements~\cite{JingGuo2020PRL}. This agreement indicates that an external field of 8~T has negligible effect on the transition temperatures of the low-temperature AFM phase. At high temperatures, our NMR measurements identify the anomaly near 160~K as a structural transition, and no NMR evidence of a phase transition near 120~K was observed. An applied field of 8~T is unlikely to qualitatively alter the high-temperature phases above 100~K. The absence of any anomaly near 120~K in our NMR data therefore suggests that the transition reported by other probes is not resolved by our local probe. Possible differences arising from pressure-cell conditions or temperature calibration by other studies require further investigation.}

{\bf Summary.}
In this work, we perform high-pressure NMR measurements on SrCu$_2$(BO$_3$)$_2$ up to 14~GPa, into a pressure regime that remains largely unexplored for strongly correlated quantum magnets. A phase diagram is established with distinct phases and features from earlier reports on SrCu$_2$(BO$_3$)$_2$. Our results reveal two structural phase transitions from the tetragonal to two monoclinic phases sequentially. The monoclinic phases host only one gapless 3D AFM ground state at low temperatures, whereas an antiferromagnetic liquid phase with short-range ordering is identified to persist above the N\'eel temperature. At even higher temperatures, a sublinear temperature dependence of the spin-lattice relaxation rate, together with the absence of NMR line split or broadening, reveals an extended correlated paramagnetic regime.

{\bf Acknowledgments.}
    This work is supported by the Scientific Research Innovation Capability Support Project for Young Faculty (Grant No. ZYGXQNJSKYCXNLZCXM-M26), the National Key Research and Development Program of China (Grant Nos. 2023YFA1406500 and 2025YFA1412100), and the National Natural Science Foundation of China (Grant Nos.~12374156 and 12134020). A portion of this work  was carried out at the Synergetic Extreme Condition User Facility (SECUF, https://cstr.cn/31123.02.SECUF.).

\bibliography{SCBOref}

@article{Balents2010Nature,
  author  = {Balents, L.},
  title   = {Spin liquids in frustrated magnets},
  journal = {Nature},
  year    = {2010},
  date    = {2010-03-01},
  volume  = {464},
  number  = {7286},
  pages   = {199--208},
  issn    = {1476-4687},
  doi     = {10.1038/nature08917},
  url     = {https://doi.org/10.1038/nature08917},
}

@article{Savary_2017,
doi = {10.1088/0034-4885/80/1/016502},
url = {https://doi.org/10.1088/0034-4885/80/1/016502},
year = {2016},
month = {nov},
publisher = {IOP Publishing},
volume = {80},
number = {1},
pages = {016502},
author = {Savary, L. and Balents, L.},
title = {Quantum spin liquids: a review},
journal = {Rep. Prog. Phys.},
}

@article{RevModPhys.89.025003,
  title = {Quantum spin liquid states},
  author = {Zhou, Y. and Kanoda, K. and Ng, T.-K.},
  journal = {Rev. Mod. Phys.},
  volume = {89},
  issue = {2},
  pages = {025003},
  numpages = {50},
  year = {2017},
  month = {Apr},
  publisher = {American Physical Society},
  doi = {10.1103/RevModPhys.89.025003},
  url = {https://link.aps.org/doi/10.1103/RevModPhys.89.025003}
}

@article{JingGuo2020PRL,
  title = {Quantum Phases of {SrCu}\(_2\)({BO}\(_3\))\(_2\) from High-Pressure Thermodynamics},
  author = {Guo, J. and Sun, G. and Zhao, B. and Wang, L. and Hong, W. and Sidorov, V. A. and Ma, N. and Wu, Q. and Li, S. and Meng, Z. and Sandvik, A. W. and Sun, L.},
  journal = {Phys. Rev. Lett.},
  volume = {124},
  issue = {20},
  pages = {206602},
  numpages = {6},
  year = {2020},
  month = {May},
  publisher = {American Physical Society},
  doi = {10.1103/PhysRevLett.124.206602},
  url = {https://link.aps.org/doi/10.1103/PhysRevLett.124.206602}
}

@article{Jimenez2021,
  author = {J. L. Jiménez and S. P. G. Crone and E. Fogh and M. E. Zayed and R. Lortz and E. Pomjakushina and  K. Conder and A. M. Läuchli and L. Weber and S. Wessel and A. Honecker and B. Normand and Ch. Rüegg and P. Corboz and H. M. Rønnow and F. Mila},
  title = {A quantum magnetic analogue to the critical point of water},
  journal = {Nature},
  year = {2021},
  volume = {592},
  number = {7854},
  pages = {370--375},
  doi = {10.1038/s41586-021-03411-8}
}

@article{Zayed2014,
title = {Temperature dependence of the pressure induced monoclinic distortion in the spin S=1/2 {Shastry–Sutherland} compound {SrCu}\(_2\)({BO}\(_3\))\(_2\)},
author = {M.E. Zayed and C. Rüegg and E. Pomjakushina and M. Stingaciu and K. Conder and M. Hanfland and M. Merlini and H.M. Rønnow},
journal = {Solid State Commun.},
volume = {186},
pages = {13-17},
year = {2014},
issn = {0038-1098},
doi = {https://doi.org/10.1016/j.ssc.2014.01.008},
url = {https://www.sciencedirect.com/science/article/pii/S0038109814000210},
}

@article{S.Haravifard2014PNAS,
author = {S. Haravifard  and A. Banerjee  and J. van Wezel  and D. M. Silevitch  and A. M. dos Santos  and J. C. Lang  and E. Kermarrec  and G. Srajer  and B. D. Gaulin  and J. J. Molaison  and H. A. Dabkowska  and T. F. Rosenbaum },
title = {Emergence of long-range order in sheets of magnetic dimers},
journal = {Proc. Natl. Acad. Sci. USA},
volume = {111},
number = {40},
pages = {14372-14377},
year = {2014},
doi = {10.1073/pnas.1413318111},
URL = {https://www.pnas.org/doi/abs/10.1073/pnas.1413318111},
}

@article{Zayed2017NaturePhysics,
    author = {M. E. Zayed and Ch. R{\"u}egg and J. {Larrea J.} and A. M. L{\"a}uchli and C. Panagopoulos and S. S. Saxena and M. Ellerby and D. F. McMorrow and Th. Strässle and S. Klotz and G. Hamel and R. A. Sadykov and V. Pomjakushin and M. Boehm and M. Jiménez–Ruiz and A. Schneidewind and E. Pomjakushina and M. Stingaciu and K. Conder and H. M. Rønnow },
    title = {4-spin plaquette singlet state in the {Shastry–Sutherland} compound {SrCu}\(_2\)({BO}\(_3\))\(_2\)},
    journal = {Nat. Phys.},
    year = {2017},
    volume = {13},
    number = {10},
    pages = {962--966},
    doi = {10.1038/nphys4190},
    url = {https://doi.org/10.1038/nphys4190}
}

@article{JingGuo2025,
    author = {Guo, J. and Wang, P. and Huang, C. and Chen, B. and Hong, W. and Cai, S. and Zhao, J. and Han, J. and Chen, X. and Zhou, Y. and Li, S. and Wu, Q. and Meng, Z. and Sun, L.},
    title = {Deconfined quantum critical point lost in pressurized {SrCu}\(_2\)({BO}\(_3\))\(_2\)},
    journal = {Commun. Phys.},
    year = {2025},
    volume = {8},
    number = {1},
    pages = {75},
    doi = {10.1038/s42005-025-01976-8},
    url = {https://doi.org/10.1038/s42005-025-01976-8}
}

@article{FoghEllen2024PRL,
  title = {Spin Waves and Three Dimensionality in the High-Pressure Antiferromagnetic Phase of {SrCu}\(_2\)({BO}\(_3\))\(_2\)},
  author = {Fogh, E. and Giriat, G. and Zayed, M. E. and Piovano, A. and Boehm, M. and Steffens, P. and Safiulina, I. and Hansen, U. B. and Klotz, S. and Soh, J. and Pomjakushina, E. and Mila, F. and Normand, B. and R\o{}nnow, H. M.},
  journal = {Phys. Rev. Lett.},
  volume = {133},
  issue = {24},
  pages = {246702},
  numpages = {7},
  year = {2024},
  month = {Dec},
  publisher = {American Physical Society},
  doi = {10.1103/PhysRevLett.133.246702},
  url = {https://link.aps.org/doi/10.1103/PhysRevLett.133.246702}
}

@article{Loa2005,
title = {Crystal structure and lattice dynamics of {SrCu}\(_2\)({BO}\(_3\))\(_2\) at high pressures},
journal = {Physica B: Condens. Matter},
volume = {359-361},
pages = {980-982},
year = {2005},
issn = {0921-4526},
doi = {https://doi.org/10.1016/j.physb.2005.01.284},
url = {https://www.sciencedirect.com/science/article/pii/S0921452605003194},
author = {I. Loa and F. Zhang and K. Syassen and P. Lemmens and W. Crichton and H. Kageyama and Y. Ueda}
}

@article{Waki2007,
author = {Waki ,T. and Arai ,K. and Takigawa ,M. and Saiga ,Y. and Uwatoko ,Y. and Kageyama ,H. and Ueda ,Y.},
title = {A Novel Ordered Phase in {SrCu}\(_2\)({BO}\(_3\))\(_2\) under High Pressure},
journal = {J. Phys. Soc. Jpn.},
volume = {76},
number = {7},
pages = {073710},
year = {2007},
doi = {10.1143/JPSJ.76.073710},
URL = {https://doi.org/10.1143/JPSJ.76.073710},
}

@article{Cui_2025,
doi = {10.1088/0256-307X/42/4/047503},
url = {https://doi.org/10.1088/0256-307X/42/4/047503},
year = {2025},
month = {apr},
publisher = {Chin. Physical Society and IOP Publishing Ltd},
volume = {42},
number = {4},
pages = {047503},
author = {Cui, Y. and Yu, R. and Yu, W.},
title = {Deconfined Quantum Critical Point: A Review of Progress},
journal = {Chin. Phys. Lett.},
}

@article{Cui2023,
author = {Y. Cui  and L. Liu  and H. Lin  and K. Wu  and W. Hong  and X. Liu  and C. Li  and Z. Hu  and N. Xi  and S. Li  and R. Yu  and A. W. Sandvik  and W. Yu },
title = {Proximate deconfined quantum critical point in  {SrCu}\(_2\)({BO}\(_3\))\(_2\)},
journal = {Science},
volume = {380},
number = {6650},
pages = {1179-1184},
year = {2023},
doi = {10.1126/science.adc9487},
URL = {https://www.science.org/doi/abs/10.1126/science.adc9487},
}

@article{SS1981,
title = {Exact ground state of a quantum mechanical antiferromagnet},
journal = {Physica B+C},
volume = {108},
number = {1},
pages = {1069-1070},
year = {1981},
issn = {0378-4363},
doi = {https://doi.org/10.1016/0378-4363(81)90838-X},
url = {https://www.sciencedirect.com/science/article/pii/037843638190838X},
author = {B. {S. Shastry} and B. Sutherland},
}

@article{Koga2000PRL,
  title = {Quantum Phase Transitions in the {Shastry-Sutherland} Model for  {SrCu}\(_2\)({BO}\(_3\))\(_2\)},
  author = {Koga, A. and Kawakami, N.},
  journal = {Phys. Rev. Lett.},
  volume = {84},
  issue = {19},
  pages = {4461--4464},
  numpages = {0},
  year = {2000},
  month = {May},
  publisher = {American Physical Society},
  doi = {10.1103/PhysRevLett.84.4461},
  url = {https://link.aps.org/doi/10.1103/PhysRevLett.84.4461}
}

@article{Manousakis1991,
  title = {The spin-1/2 Heisenberg antiferromagnet on a square lattice and its application to the cuprous oxides},
  author = {Manousakis, E.},
  journal = {Rev. Mod. Phys.},
  volume = {63},
  issue = {1},
  pages = {1--62},
  numpages = {0},
  year = {1991},
  month = {Jan},
  publisher = {American Physical Society},
  doi = {10.1103/RevModPhys.63.1},
  url = {https://link.aps.org/doi/10.1103/RevModPhys.63.1}
}

@article{Corboz2013PRB,
  title = {Tensor network study of the {Shastry-Sutherland} model in zero magnetic field},
  author = {Corboz, P. and Mila, F.},
  journal = {Phys. Rev. B},
  volume = {87},
  issue = {11},
  pages = {115144},
  numpages = {10},
  year = {2013},
  month = {Mar},
  publisher = {American Physical Society},
  doi = {10.1103/PhysRevB.87.115144},
  url = {https://link.aps.org/doi/10.1103/PhysRevB.87.115144}
}

@article{Sakurai2018,
author = {Sakurai ,T. and Hirao ,Y. and Hijii ,K. and Okubo ,S. and Ohta ,H. and Uwatoko ,Y. and Kudo ,K. and Koike ,Y.},
title = {Direct Observation of the Quantum Phase Transition of {SrCu}\(_2\)({BO}\(_3\))\(_2\) by High-Pressure and Terahertz Electron Spin Resonance},
journal = {J. Phys. Soc. Jpn.},
volume = {87},
number = {3},
pages = {033701},
year = {2018},
doi = {10.7566/JPSJ.87.033701},
URL = { https://doi.org/10.7566/JPSJ.87.033701},
}

@article{Miyahara1999PRL,
  title = {Exact Dimer Ground State of the Two Dimensional Heisenberg Spin System  {SrCu}\(_2\)({BO}\(_3\))\(_2\)},
  author = {Miyahara, S. and Ueda, K.},
  journal = {Phys. Rev. Lett.},
  volume = {82},
  issue = {18},
  pages = {3701--3704},
  numpages = {0},
  year = {1999},
  month = {May},
  publisher = {American Physical Society},
  doi = {10.1103/PhysRevLett.82.3701},
  url = {https://link.aps.org/doi/10.1103/PhysRevLett.82.3701}
}

@article{Hong-Qiang_1991_PRB,
  title = {Two-dimensional spin-1/2 Heisenberg antiferromagnet: A quantum Monte Carlo study},
  author = {Makivi\ifmmode \acute{c}\else \'{c}\fi{}, M. S. and Ding, H.},
  journal = {Phys. Rev. B},
  volume = {43},
  issue = {4},
  pages = {3562--3574},
  numpages = {0},
  year = {1991},
  month = {Feb},
  publisher = {American Physical Society},
  doi = {10.1103/PhysRevB.43.3562},
  url = {https://link.aps.org/doi/10.1103/PhysRevB.43.3562}
}

@article{Gomez-Santos_1989_PRB,
  title = {Monte Carlo study of the quantum spin-1/2 Heisenberg antiferromagnet on the square lattice},
  author = {Gomez-Santos, G. and Joannopoulos, J. D. and Negele, J. W.},
  journal = {Phys. Rev. B},
  volume = {39},
  issue = {7},
  pages = {4435--4443},
  numpages = {0},
  year = {1989},
  month = {Mar},
  publisher = {American Physical Society},
  doi = {10.1103/PhysRevB.39.4435},
  url = {https://link.aps.org/doi/10.1103/PhysRevB.39.4435}
}

@article{Kim_1998_PRL,
  title = {Low Temperature Behavior and Crossovers of the Square Lattice Quantum Heisenberg Antiferromagnet},
  author = {Kim, J.-K. and Troyer, M.},
  journal = {Phys. Rev. Lett.},
  volume = {80},
  issue = {12},
  pages = {2705--2708},
  numpages = {0},
  year = {1998},
  month = {Mar},
  publisher = {American Physical Society},
  doi = {10.1103/PhysRevLett.80.2705},
  url = {https://link.aps.org/doi/10.1103/PhysRevLett.80.2705}
}

@article{Kageyama_1999_JPSJ,
author = {Kageyama ,H. and Onizuka ,K. and Yamauchi ,T. and Ueda ,Y. and Hane ,S. and Mitamura ,H. and Goto ,T. and Yoshimura ,K. and Kosuge ,K.},
title = {Anomalous Magnetizations in Single Crystalline {SrCu}\(_2\)({BO}\(_3\))\(_2\)},
journal = {J. Phys. Soc. Jpn.},
volume = {68},
number = {6},
pages = {1821-1823},
year = {1999},
doi = {10.1143/JPSJ.68.1821},
URL = { https://doi.org/10.1143/JPSJ.68.1821},

}

@article{Shi2022NC,
  author    = {Z. Shi and S. Dissanayake and P. Corboz and W. Steinhardt and D. Graf and D. M. Silevitch and A. Hanna. Dabkowska and T. F. Rosenbaum and F. Mila and S. Haravifard},
  title     = {Discovery of quantum phases in the {Shastry-Sutherland} compound {SrCu$_2$(BO$_3$)$_2$} under extreme conditions of field and pressure},
  journal   = {Nat. Commun.},
  year      = {2022},
  volume    = {13},
  number    = {1},
  pages     = {2301},
  month     = {apr},
  date      = {2022-04-28},
  doi       = {10.1038/s41467-022-30036-w},
  url       = {https://doi.org/10.1038/s41467-022-30036-w},
}

@article{Chen2018ScienceBulletin,
title = {Thermodynamics of spin-1/2 Kagomé Heisenberg antiferromagnet: algebraic paramagnetic liquid and finite-temperature phase diagram},
journal = {Sci. Bull.},
volume = {63},
number = {23},
pages = {1545-1550},
year = {2018},
issn = {2095-9273},
doi = {https://doi.org/10.1016/j.scib.2018.11.007},
url = {https://www.sciencedirect.com/science/article/pii/S2095927318305267},
author = {X. Chen and S. Ran and T. Liu and C. Peng and Y. Huang and G. Su},
}

@article{17OPRL,
  title = {Field-Induced Freezing of a Quantum Spin Liquid on the Kagome Lattice},
  author = {Jeong, M. and Bert, F. and Mendels, P. and Duc, F. and Trombe, J. C. and de Vries, M. A. and Harrison, A.},
  journal = {Phys. Rev. Lett.},
  volume = {107},
  issue = {23},
  pages = {237201},
  numpages = {5},
  year = {2011},
  month = {Nov},
  publisher = {American Physical Society},
  doi = {10.1103/PhysRevLett.107.237201},
  url = {https://link.aps.org/doi/10.1103/PhysRevLett.107.237201}
}

@article{XiNing2023PRB,
  title = {Plaquette valence bond solid to antiferromagnet transition and deconfined quantum critical point of the {Shastry-Sutherland} model},
  author = {Xi, N. and Chen, H. and Xie, Z. Y. and Yu, R.},
  journal = {Phys. Rev. B},
  volume = {107},
  issue = {22},
  pages = {L220408},
  numpages = {6},
  year = {2023},
  month = {Jun},
  publisher = {American Physical Society},
  doi = {10.1103/PhysRevB.107.L220408},
  url = {https://link.aps.org/doi/10.1103/PhysRevB.107.L220408}
}

@Article{Sandvik2022CPL,
title = {Quantum Spin Liquid Phase in the {Shastry–Sutherland} Model Detected by an Improved Level Spectroscopic Method},
journal = {Chin. Phys. Lett.},
volume = {39},
number = {7},
pages = {077502-077502},
year = {2022},
issn = {},
doi = {10.1088/0256-307X/39/7/077502},	
url = {http://cpl.iphy.ac.cn/en/article/doi/10.1088/0256-307X/39/7/077502},
author = {L. Wang and Y. Zhang and A. W. Sandvik}
}

@article{Wangling2022PRB,
  title = {Quantum criticality and spin liquid phase in the {Shastry-Sutherland} model},
  author = {Yang, J. and Sandvik, A. W. and Wang, L.},
  journal = {Phys. Rev. B},
  volume = {105},
  issue = {6},
  pages = {L060409},
  numpages = {7},
  year = {2022},
  month = {Feb},
  publisher = {American Physical Society},
  doi = {10.1103/PhysRevB.105.L060409},
  url = {https://link.aps.org/doi/10.1103/PhysRevB.105.L060409}
}

@article{Viteritti2025PRB,
  title = {Transformer wave function for two dimensional frustrated magnets: Emergence of a spin-liquid phase in the {Shastry-Sutherland} model},
  author = {Viteritti, L. L. and Rende, R. and Parola, A. and Goldt, S. and Becca, F.},
  journal = {Phys. Rev. B},
  volume = {111},
  issue = {13},
  pages = {134411},
  numpages = {15},
  year = {2025},
  month = {Apr},
  publisher = {American Physical Society},
  doi = {10.1103/PhysRevB.111.134411},
  url = {https://link.aps.org/doi/10.1103/PhysRevB.111.134411}
}

@misc{Mengziyang2026arxiv,
      title={Thermodynamics of {Shastry-Sutherland} Model under Magnetic Field}, 
      author={M. Song and C. Zhou and C. Huang and Z. Meng},
      eprint={2602.11589},
      archivePrefix={arXiv},
}

@misc{Guo2026arxiv,
      title={T-linear specific heat in pressurized and magnetized {Shastry-Sutherland} Mott insulator {SrCu}\(_2\)({BO}\(_3\))\(_2\)}, 
      author={J. Guo and P. Wang and C. Huang and C. Zhou and M. Song and X. Chen and T. Wang and W. Hong and S. Cai and J. Zhao and J. Han and Y. Zhou and Q. Wu and S. Li and Z. Meng and L. Sun},
      eprint={2602.18229},
      archivePrefix={arXiv},
}

@article{KAGEYAMA199965,
title = {Crystal growth of the two-dimensional spin gap system {SrCu}\(_2\)({BO}\(_3\))\(_2\)},
journal = {J. Cryst. Growth},
volume = {206},
number = {1},
pages = {65-67},
year = {1999},
issn = {0022-0248},
doi = {https://doi.org/10.1016/S0022-0248(99)00313-9},
url = {https://www.sciencedirect.com/science/article/pii/S0022024899003139},
author = {H. Kageyama and K. Onizuka and T. Yamauchi and Y. Ueda},
}

@article{K.Onizuka2000JPSJ,
author = {Onizuka ,K. and Kageyama ,H. and Narumi ,Y. and Kindo ,K. and Ueda ,Y. and Goto ,T.},
title = {1/3 Magnetization Plateau in {SrCu}\(_2\)({BO}\(_3\))\(_2\) - Stripe Order of Excited Triplets},
journal = {J. Phys. Soc. Jpn.},
volume = {69},
number = {4},
pages = {1016-1018},
year = {2000},
doi = {10.1143/JPSJ.69.1016},

URL = {https://doi.org/10.1143/JPSJ.69.1016},
}

@article{D.A.Schneider2016PRB,
  title = {Pressure dependence of the magnetization plateaus of {SrCu}\(_2\)({BO}\(_3\))\(_2\)},
  author = {Schneider, D. A. and Coester, K. and Mila, F. and Schmidt, K. P.},
  journal = {Phys. Rev. B},
  volume = {93},
  issue = {24},
  pages = {241107},
  numpages = {5},
  year = {2016},
  month = {Jun},
  publisher = {American Physical Society},
  doi = {10.1103/PhysRevB.93.241107},
  url = {https://link.aps.org/doi/10.1103/PhysRevB.93.241107}
}

@article{WOLF20011973,
title = {Ultrasonic experiments in {SrCu}\(_2\)({BO}\(_3\))\(_2\) and  {NH}\(_3\){CuCl}\(_3\) in magnetic fields up to 50{T}},
journal = {J. Magn. Magn. Mater.},
volume = {226-230},
pages = {1973-1975},
year = {2001},
issn = {0304-8853},
doi = {https://doi.org/10.1016/S0304-8853(01)00093-2},
url = {https://www.sciencedirect.com/science/article/pii/S0304885301000932},
author = {B. Wolf and S. Zherlitsyn and S. Schmidt and H. Schwenk and B. Lüthi and H. Kageyama and Y. Ueda and H. Tanaka},
}

@article{Y.Fukumoto2001JPSJ,
author = {Fukumoto ,Y.},
title = {Magnetization Plateaus in the {Shastry-Sutherland} Model for {SrCu}\(_2\)({BO}\(_3\))\(_2\): Results of Fourth-Order Perturbation Expansion with a Low-Density Approximation},
journal = {J. Phys. Soc. Jpn.},
volume = {70},
number = {5},
pages = {1397-1403},
year = {2001},
doi = {10.1143/JPSJ.70.1397},
URL = { https://doi.org/10.1143/JPSJ.70.1397},
}

@article{S.Miyahara2000PRB,
  title = {Superstructures at magnetization plateaus in {SrCu}\(_2\)({BO}\(_3\))\(_2\)},
  author = {Miyahara, S. and Ueda, K.},
  journal = {Phys. Rev. B},
  volume = {61},
  issue = {5},
  pages = {3417--3424},
  numpages = {0},
  year = {2000},
  month = {Feb},
  publisher = {American Physical Society},
  doi = {10.1103/PhysRevB.61.3417},
}

@article{Fogh2024,
  author = {Fogh, E. and Nayak, M. and Prokhnenko, O. and Bartkowiak, M. and Munakata, K. and Soh, J.-R. and A. A. Turrini and M.E. Zayed and E. Pomjakushina and H. Kageyama and H. Nojiri and K. Kakurai and B. Normand and F. Mila and H. M. Rønnow },
  title = {Field-induced bound-state condensation and spin-nematic phase in {SrCu}\(_2\)({BO}\(_3\))\(_2\) revealed by neutron scattering up to 25.9~{T} },
  journal = {Nat. Commun.},
  year = {2024},
  volume = {15},
  number = {1},
  pages = {442},
  doi = {10.1038/s41467-023-44115-z},
  url = {https://doi.org/10.1038/s41467-023-44115-z}
}

@article{S.Imajo2022PRL,
  title = {Magnetically Hidden State on the Ground Floor of the Magnetic Devil's Staircase},
  author = {Imajo, S. and Matsuyama, N. and Nomura, T. and Kihara, T. and Nakamura, S. and Marcenat, C. and Klein, T. and Seyfarth, G. and Zhong, C. and Kageyama, H. and Kindo, K. and Momoi, T. and Kohama, Y.},
  journal = {Phys. Rev. Lett.},
  volume = {129},
  issue = {14},
  pages = {147201},
  numpages = {6},
  year = {2022},
  month = {Sep},
  publisher = {American Physical Society},
  doi = {10.1103/PhysRevLett.129.147201},
  url = {https://link.aps.org/doi/10.1103/PhysRevLett.129.147201}
}

@article{S.C.Furuya2018PRB,
  title = {Electron spin resonance for the detection of long-range spin nematic order},
  author = {Furuya, S. C. and Momoi, T.},
  journal = {Phys. Rev. B},
  volume = {97},
  issue = {10},
  pages = {104411},
  numpages = {18},
  year = {2018},
  month = {Mar},
  publisher = {American Physical Society},
  doi = {10.1103/PhysRevB.97.104411},
  url = {https://link.aps.org/doi/10.1103/PhysRevB.97.104411}
}

@article{T.Momoi2000PRB,
  title = {Magnetization plateaus of the {Shastry-Sutherland} model for {SrCu}\(_2\)({BO}\(_3\))\(_2\): Spin-density wave, supersolid, and bound states},
  author = {Momoi, T. and Totsuka, K.},
  journal = {Phys. Rev. B},
  volume = {62},
  issue = {22},
  pages = {15067--15078},
  numpages = {0},
  year = {2000},
  month = {Dec},
  publisher = {American Physical Society},
  doi = {10.1103/PhysRevB.62.15067},
  url = {https://link.aps.org/doi/10.1103/PhysRevB.62.15067}
}

@article{T.Nomura2023NC,
  author = {Nomura, T. and Corboz, P. and Miyata, A. and Zherlitsyn, S. and Ishii, Y. and Kohama, Y. and Y. H. Matsuda and A. Ikeda and C. Zhong and H. Kageyama and F. Mila },
  title = {Unveiling new quantum phases in the {Shastry-Sutherland} compound {SrCu}\(_2\)({BO}\(_3\))\(_2\) up to the saturation magnetic field},
  journal = {Nat. Commun.},
  year = {2023},
  volume = {14},
  number = {1},
  pages = {3769},
  doi = {10.1038/s41467-023-39502-5},
  url = {https://doi.org/10.1038/s41467-023-39502-5}
}

@article{P.Corboz2014PRL,
  title = {Crystals of Bound States in the Magnetization Plateaus of the {Shastry-Sutherland} Model},
  author = {Corboz, P. and Mila, F.},
  journal = {Phys. Rev. Lett.},
  volume = {112},
  issue = {14},
  pages = {147203},
  numpages = {5},
  year = {2014},
  month = {Apr},
  publisher = {American Physical Society},
  doi = {10.1103/PhysRevLett.112.147203},
  url = {https://link.aps.org/doi/10.1103/PhysRevLett.112.147203}
}

@article{Mila2026PRL,
  title = {Quantum Spin Liquid Phase in the {S}hastry-{S}utherland Model Revealed by High-Precision Infinite Projected Entangled-Pair States},
  author = {P. Corboz and Y. Zhang and B. Ponsioen and F. Mila},
  journal = {Phys. Rev. Lett.},
  volume = {136},
  issue = {18},
  pages = {186701},
  numpages = {8},
  year = {2026},
  month = {May},
  publisher = {American Physical Society},
  doi = {10.1103/67xm-5d8v},
  url = {https://link.aps.org/doi/10.1103/67xm-5d8v}
}

@article{Cava2010PRB,
  title = {Coexisting magnetic order and cooperative paramagnetism in the stuffed pyrochlore {Tb}\(_{2+x}\){Ti}\(_{2-2x}\){Nb}\(_{x}\){O}\(_7\)},
  author = {Ueland, B. G. and Gardner, J. S. and Williams, A. J. and Dahlberg, M. L. and Kim, J. G. and Qiu, Y. and Copley, J. R. D. and Schiffer, P. and Cava, R. J.},
  journal = {Phys. Rev. B},
  volume = {81},
  issue = {6},
  pages = {060408(R)},
  numpages = {4},
  year = {2010},
  month = {Feb},
  publisher = {American Physical Society},
  doi = {10.1103/PhysRevB.81.060408},
  url = {https://link.aps.org/doi/10.1103/PhysRevB.81.060408}
}

@article{Ruiz2017NC,
  author = {Ruiz, A. and Frano, A. and Breznay, N. P. and Kimchi, I. and Helm, T. and Oswald, I. and Chan, J. Y. and Birgeneau, R. J. and Islam, Z. and Analytis, J. G.},
  title  = {{Correlated states in $\beta$-Li$_2$IrO$_3$ driven by applied magnetic fields}},
  journal = {Nat. Commun.},
  volume = {8},
  number = {1},
  pages  = {961},
  year   = {2017},
  doi    = {10.1038/s41467-017-01071-9},
  issn   = {2041-1723},
  url    = {https://doi.org/10.1038/s41467-017-01071-9}
}

@article{Singh_2019,
doi = {10.1088/1361-648X/ab3c14},
url = {https://doi.org/10.1088/1361-648X/ab3c14},
year = {2019},
month = {sep},
publisher = {IOP Publishing},
volume = {31},
number = {48},
pages = {485803},
author = {Singh, B. and Kumar, D. and Manna, K. and Bera, A. K. and Cansever, G. A. and Maljuk, A. and Wurmehl, S. and Büchner, B. and Kumar, P.},
title = {Correlated paramagnetism and interplay of magnetic and phononic degrees of freedom in 3d-5d coupled {La}\(_2\){Cu}{Ir}{O}\(_2\)},
journal = {J. Phys.: Condens. Matter},
}

@article{Cederholm2026,
author = {Cederholm, J. J. and Piovano, A. and Ivanov, A. and Klotz, S. and Hiess, A. and Hansen, U. B. and Pomjakushina, E. and Zayed, M. E. and Fogh, E. and R{\o}nnow, H. M.},
title = {Spin excitations near the pressure-induced antiferromagnetic transition in {SrCu}\(_2\)({BO}\(_3\))\(_2\)},
journal = {J. Appl. Crystallogr.},
year = {2026},
volume = {59},
number = {3},
pages = "",
month = {Jun},
url = {https://doi.org/10.1107/S1600576726003705},
}

@article{Haravifard2012PNAS,
author = {S. Haravifard  and A. Banerjee  and J. C. Lang  and G. Srajer  and D. M. Silevitch  and B. D. Gaulin  and H. A. Dabkowska  and T. F. Rosenbaum },
title = {Continuous and discontinuous quantum phase transitions in a model two-dimensional magnet},
journal = {Proc. Natl. Acad. Sci. USA},
volume = {109},
number = {7},
pages = {2286-2289},
year = {2012},
doi = {10.1073/pnas.1114464109},
URL = {https://www.pnas.org/doi/abs/10.1073/pnas.1114464109},
}

@article{2018PRX,
  title = {Signatures of a Deconfined Phase Transition on the {Shastry-Sutherland} Lattice: Applications to Quantum Critical {SrCu}\(_2\)({BO}\(_3\))\(_2\)},
  author = {Lee, J. and You, Y. and Sachdev, S. and Vishwanath, A.},
  journal = {Phys. Rev. X},
  volume = {9},
  issue = {4},
  pages = {041037},
  numpages = {25},
  year = {2019},
  month = {Nov},
  publisher = {American Physical Society},
  doi = {10.1103/PhysRevX.9.041037},
  url = {https://link.aps.org/doi/10.1103/PhysRevX.9.041037}
}

@article{Albert1967,
  title = {Nuclear Spin-Lattice Relaxation in Hexagonal Transition Metals: Titanium},
  author = {Narath, A.},
  journal = {Phys. Rev.},
  volume = {162},
  issue = {2},
  pages = {320--332},
  numpages = {0},
  year = {1967},
  month = {Oct},
  publisher = {American Physical Society},
  doi = {10.1103/PhysRev.162.320},
  url = {https://link.aps.org/doi/10.1103/PhysRev.162.320}
}

@article{Chepin_1991,
doi = {10.1088/0953-8984/3/41/009},
url = {https://doi.org/10.1088/0953-8984/3/41/009},
year = {1991},
month = {oct},
publisher = {},
volume = {3},
number = {41},
pages = {8103},
author = {Chepin, J. and Ross, J. H., Jr.},
title = {Magnetic spin-lattice relaxation in nuclear quadrupole resonance: the $\eta$ $\neq$ 0 case},
journal = {J. Phys.: Condens. Matter},
}

@article{Khuntia2016PRL,
  title = {Spin Liquid State in the 3{D} Frustrated Antiferromagnet {${\mathrm{PbCuTe}}_{2}{\mathrm{O}}_{6}$}: {NMR} and Muon Spin Relaxation Studies},
  author = {Khuntia, P. and Bert, F. and Mendels, P. and Koteswararao, B. and Mahajan, A. V. and Baenitz, M. and Chou, F. C. and Baines, C. and Amato, A. and Furukawa, Y.},
  journal = {Phys. Rev. Lett.},
  volume = {116},
  issue = {10},
  pages = {107203},
  numpages = {5},
  year = {2016},
  month = {Mar},
  publisher = {American Physical Society},
  doi = {10.1103/PhysRevLett.116.107203},
  url = {https://link.aps.org/doi/10.1103/PhysRevLett.116.107203}
}

@article{Knafo2017PRB,
  title = {Three-dimensional critical phase diagram of the Ising antiferromagnet {${\mathrm{CeRh}}_{2}{\mathrm{Si}}_{2}$} under intense magnetic field and pressure},
  author = {Knafo, W. and Settai, R. and Braithwaite, D. and Kurahashi, S. and Aoki, D. and Flouquet, J.},
  journal = {Phys. Rev. B},
  volume = {95},
  issue = {1},
  pages = {014411},
  numpages = {10},
  year = {2017},
  month = {Jan},
  publisher = {American Physical Society},
  doi = {10.1103/PhysRevB.95.014411},
  url = {https://link.aps.org/doi/10.1103/PhysRevB.95.014411}
}

@article{Posp2018PRB,
  title = {Magnetic field induced phenomena in UIrGe in fields applied along the $b$ axis},
  author = {Pospíšil, J. and Haga, Y. and Kohama, Y. and Miyake, A. and Kambe, S. and Tateiwa, N. and Vališka, M. and Proschek, P. and Prokleška, J. and Sechovský, V. and Tokunaga, M. and Kindo, K. and Matsuo, A. and Yamamoto, E.},
  journal = {Phys. Rev. B},
  volume = {98},
  issue = {1},
  pages = {014430},
  numpages = {7},
  year = {2018},
  month = {Jul},
  publisher = {American Physical Society},
  doi = {10.1103/PhysRevB.98.014430},
  url = {https://link.aps.org/doi/10.1103/PhysRevB.98.014430}
}

@article{Lee2016PRB,
  title = {Multistage symmetry breaking in the breathing pyrochlore lattice {${{\mathrm{Li(Ga,In)}}{\mathrm{Cr}}_{4}{\mathrm{O}}_{8}}$}},
  author = {Lee, S. and Do, S.-H. and Lee, W.-J. and Choi, Y. S. and Lee, M. and Choi, E. S. and Reyes, A. P. and Kuhns, P. L. and Ozarowski, A. and Choi, K.-Y.},
  journal = {Phys. Rev. B},
  volume = {93},
  pages = {174402},
  year = {2016},
  doi = {10.1103/PhysRevB.93.174402},
  url = {https://link.aps.org/doi/10.1103/PhysRevB.93.174402}
}

@article{Graham2023PRL,
  title = {Experimental Evidence for the Spiral Spin Liquid in {LiYbO$_2$}},
  author = {Graham, J. N. and Qureshi, N. and Ritter, C. and Manuel, P. and Wildes, A. R. and Clark, L.},
  journal = {Phys. Rev. Lett.},
  volume = {130},
  pages = {166703},
  year = {2023},
  doi = {10.1103/PhysRevLett.130.166703},
  url = {https://link.aps.org/doi/10.1103/PhysRevLett.130.166703}
}

@article{Hovancik2022JSSCh,
  title = {{{${\mathrm{UI}_{3}}$}} – 5f-electron magnetic van der Waals material},
  author = {Hovančík, D. and Kratochvílová, M. and Doležal, P. and Bendová, A. and Pospíšil, J. and Sechovský, V.},
  journal = {J. Solid State Chem.},
  volume = {316},
  pages = {123580},
  year = {2022},
  doi = {10.1016/j.jssc.2022.123580},
  url = {https://doi.org/10.1016/j.jssc.2022.123580}
}

@article{Chaix2026GeFe2O4,
  author  = {Chaix, L. and Robert, J. and Chan, E. and Ressouche, E. and Petit, S. and Colin, C. V. and Ballou, R. and Ollivier, J. and Regnault, L.-P. and Lhotel, E. and Cathelin, V. and Lenne, S. and Cavenel, C. and Damay, F. and Suard, E. and Strobel, P. and Darie, C. and de Brion, S. and Simonet, V.},
  title   = {From Triangular Correlated Paramagnet to Multi-$q$ Noncoplanar Spin State in Spinel {GeFe$_2$O$_4$}},
  journal = {Phys. Rev. Lett.},
  volume  = {136},
  number  = {1},
  pages   = {016703},
  year    = {2026},
  doi     = {10.1103/jl6t-ymd1}
}

\end{document}